\documentclass[12pt]{article}
\usepackage{graphicx}
\usepackage{amssymb}
\usepackage{epstopdf}
\title{Accretion Disks in Large-scale Magnetic Fields}
\author{ Jean-Pierre De Villiers\footnote{current contact: jpd\_asph@mac.com}\\
\footnotesize{University of Calgary, Physics \& Astronomy, Calgary, Alberta}}

\begin{document}
\maketitle
\normalsize

\abstract{
This paper presents a survey of 
a set of simulations of accretion disks orbiting a rapidly rotating Kerr black hole and embedded in a large-scale
initial magnetic field. Each simulation uses a common state for the
initial torus, including an MRI seed-field consisting of poloidal loops
along isodensity contours, and differ only in the strength of the initial large-scale magnetic field in relation to these poloidal loops. Simulations with a 
weak large-scale initial field differ little from simulations that 
evolve from only the poloidal field. Simulations where the large-scale initial field distorts or completely overwhelms the poloidal loops show more extensive regions
of turbulence, due to the action of the MRI on the large-scale field. However,
the overall structure of the late-time state of the simulations remains
qualitatively unchanged, that is each simulation sees the emergence of a
turbulent accretion disk, axial jets, a funnel-wall outflow and an extensive corona. 
}
\section{Introduction}
\parskip=12pt

The numerical study of black hole accretion has been an area of ongoing 
research, beginning with the pioneering studies of Wilson (1972). Simulations 
of accretion driven by the Magneto-Rotational Instability (MRI; Balbus \& 
Hawley, 1998) also have a long history, with various approximations used to 
mimic the properties of the central mass (e.g., the pseudo-Newtonian 
simulations of Hawley \& Krolik, 2001). More recently, codes such as the 
General Relativistic MHD (GRMHD) 
code of De Villiers \& Hawley (2003; DH03a) have been used to study
accretion disks orbiting Kerr black holes and the jets that arise 
self-consistently from the accretion flow (De Villiers, Hawley \& Krolik, 2003; 
DHK). The key result from these simulations is that GRMHD is 
required to correctly capture the effect of black hole spin on the energy of
the jets (De Villiers et al., 2005; D05a), which are powered by 
an MHD interaction between the accretion flow and the spinning black hole. 

Another line of inquiry deals with simulations of radial accretion onto black 
holes. In such simulations, for instance those described by Komissarov (2005), an
initial state consisting of a tenuous plasma embedded in a large-scale
magnetic field is evolved. Such simulations feature many interesting aspects, of immediate interest are
the winding of the magnetic field by black hole rotation, and the generation
of broad, energetic outflows. Such simulations seem to encounter a generic problem: the outflow which develops tends to be uncollimated and severely limits further accretion since any infalling gas at large radii is soon blown out by the outflow produced near the black hole. 

Accretion disk simulations do not suffer from this drawback: the initial torus 
in the equatorial plane serves as a long-lived mass reservoir that can sustain axial outflows for very long periods (D05a). However, most of the accretion disk 
simulations carried out with the GRMHD code 
have not featured a large-scale initial magnetic field, and instead trigger the MRI in
an initial torus by using a weak initial magnetic field confined to the 
region immediately surrounding the pressure maximum of the
torus. Since ambient
fields are likely to exist in some form in an astrophysical context, it is not simply of 
academic interest to consider how the evolution of an accretion disk would be 
affected if the initial torus were immersed in a large-scale weak magnetic field. 
In the context of GRMHD simulations, this is readily achieved 
using analytic solutions for a large-scale field due to Wald (1974; W74).

Attempts at merging the two approaches with a general relativistic
code have been reported. Nishikawa {\it et al.} (2005) evolve a thin
accretion disk in the Schwarzschild metric in three spatial dimensions. However, 
those simulations use an extremely strong initial magnetic field ($\beta$ of 
order unity in the initial disk), which may pose 
problems for the growth of the MRI (Hawley \& Balbus, 1992; HB92), and are of very short duration, on the order 
of a few percent of the typical duration reported here, again precluding proper 
development of the MRI. De Villiers {\it et al.} (2005; D05b) reported 
three axisymmetric simulations and one 3D simulation of an accretion disk 
orbiting a Kerr black hole evolved from
an initial torus immersed in a weak large-scale field; the $R_{\rm vf}$
simulations, though applied to the collapsar scenario, are analogous to the weak-field simulation (the $W$ simulation; see below) discussed here.

This paper presents a family of axisymmetric (2.5D) simulations which evolve
an initial torus immersed in a Wald magnetic field and orbiting a Kerr black hole. The simulations
differ only in the relative strength of the large-scale field that is superimposed on the traditional MRI seed field, that of poloidal loops embedded in the initial torus. Technical details are left to the appendices.

\section{Overview of the Simulations}

To simulate accretion disks, begin with an initial equilibrium torus in the 
space-time of a rotating black hole. This equilibrium torus is a differentially
rotating system, with a near-Keplerian distribution of angular momentum. In 
order for this torus to evolve into an accretion disk, it needs to be perturbed 
from its equilibrium state by introducing small density 
fluctuations in the torus, and, most importantly, by supplying a weak magnetic 
field that constitutes the seed field which will trigger the Balbus-Hawley 
magneto-rotational instability. The MRI amplifies the seed magnetic field
by shearing. The MRI is characterized by a linear growth phase of the 
low-order modes of the magnetic field, a process which take place
on the time scale of the orbital period of the fluid. This is followed by an
exponential growth phase due to non-linear interaction between the growing modes. This non-linear interaction gives rise to the turbulent 
transport of angular momentum, transforming the initial torus into an 
accretion disk. In axisymmetry, the process of turbulent transport of angular momentum is not self-sustaining, due
to Cowling's anti-dynamo theorem; axisymmetric simulations have shown that
turbulence dies down, typically within a few orbits of onset.

In earlier 3D simulations (e.g. DHK), the MRI seed field was confined 
to the region near the pressure maximum of the initial torus. Two configurations have been used: poloidal loops laid 
down along isodensity contours and, less frequently, large-scale toroidal loops. 
In such simulations, the MRI evolves on the {\it local} time-scale 
dictated by the orbital period at the initial pressure maximum, since this is 
the early locus of the MRI. Therefore, the orbital period at the initial 
pressure maximum constitutes a natural measure of simulation time for these simulations. In the 2.5D 
simulations discussed here, the MRI seed field consists of poloidal loops on 
which are superimposed a pervasive large-scale field whose overall orientation 
is parallel to the spin-axis of the black hole. Since this 
large-scale field threads the entire initial torus, the orbital period at the 
pressure maximum is not used as a time measure; comparisons between
simulations will be done using the common relativistic time unit of black hole 
mass.

\begin{figure}[htbp]
\begin{center}
\includegraphics[width=6.5in]{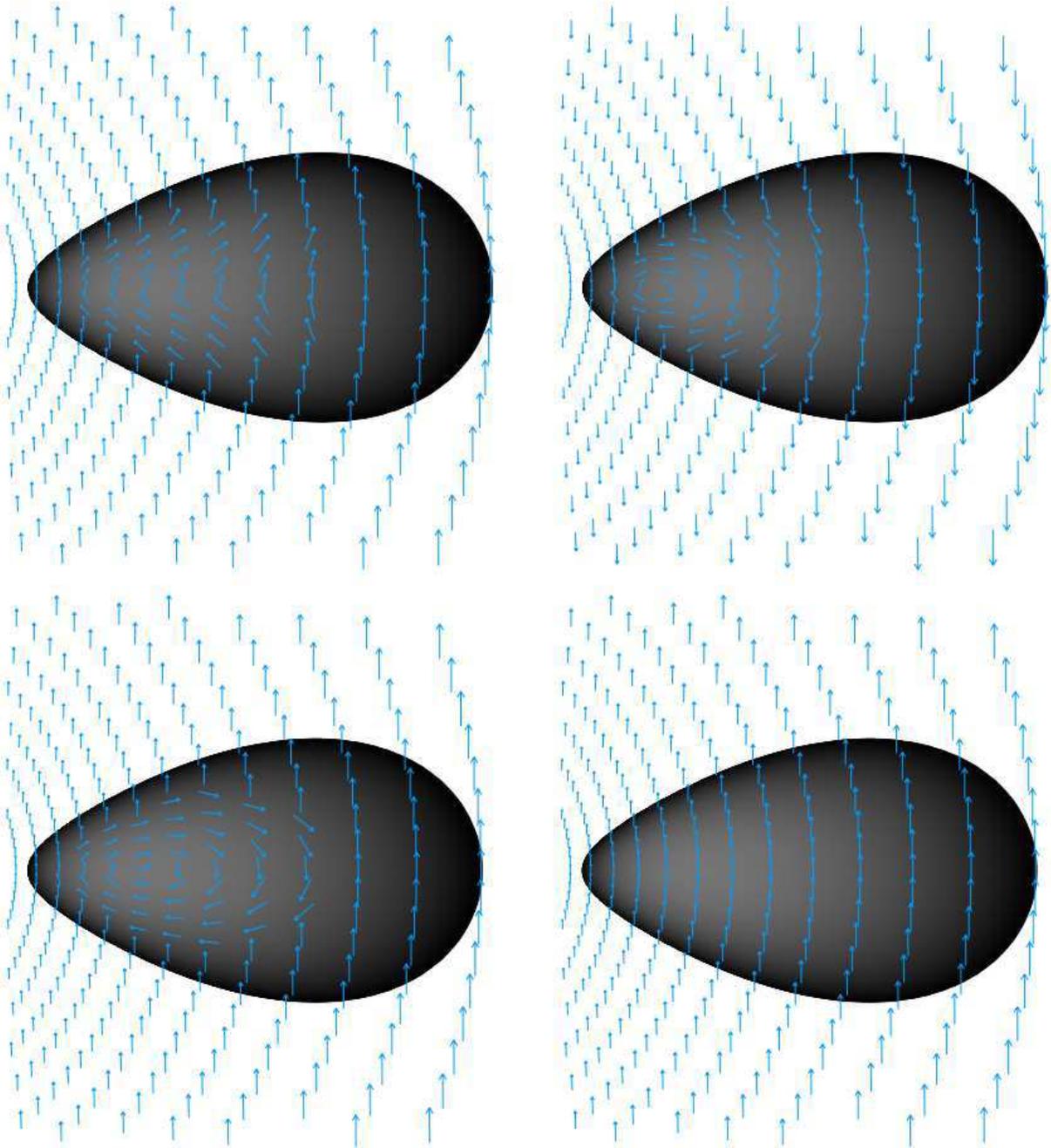}
\caption{{\small Initial state for the four simulations. The initial torus is plotted using the logarithm of density (grey scale). Arrows show the local
direction of the initial magnetic field, which is purely
poloidal. The initial states are, clockwise from bottow left: W, $M_u$, $M_d$, S.}}
\label{InitState}
\end{center}
\end{figure}

Four simulations are discussed in this paper. The simulations differ only in the 
strength and orientation (parallel or anti-parallel to the spin axis) of the large-scale 
initial magnetic field. Each simulation features a 
torus with structural parameters listed in Table \ref{params}. The initial tori are immersed in a tenuous medium which consists of a cold, radially infalling 
dust. The initial pressure maximum of these tori is at a radius or $r/M = 16.1$ in the equatorial plane, which is relatively close to the black hole; the main reason for this choice is that it allows rapid evolution of the accretion
disk. The initial tori are 
perturbed by introducing random density fluctuations ($1\%$ amplitude) as well
as a weak magnetic field in the form of poloidal loops with an average $\beta$ 
(the ratio of gas to magnetic pressure) of 100. The initial magnetic field is then 
completed by superimposing a large-scale magnetic field on the poloidal 
loops; the strength of the large-scale field is set by a scaling constant, 
$B_0$ in Wald's solution, and differs for each model. The strength of the 
large-scale field is measured in relation to the initial poloidal loops:
\begin{itemize}
\item {\bf Weak field:} where the large-scale field is much weaker in the torus than the poloidal loops and barely perturbs the structure of the loops (this simulation is denoted $W$); 
\item {\bf Moderate field:} where the large-scale field is comparable in magnitude
through the torus to the poloidal loops; two cases are considered, where the
field is oriented in the ``up'' direction ($M_u$), and the ``down'' direction ($M_d$), and depending on the orientation, this field cancels the
``vertical'' portion of the poloidal loops either inward or outward of the 
pressure maximum;
\item {\bf Strong field:} where the large-scale field dominates the poloidal loops,
effectively washing out these loops in the torus (denoted $S$). The term ``strong field'' is
a relative one; the strength of this field was adjusted to just wash out the initial poloidal
loops, and the overall strength of the field, as measured by the plasma 
$\beta$-parameter, (the ratio of gas to magnetic pressure) remains
weak.
\end{itemize}
Each of these initial states is shown in Figure \ref{InitState}, where gas density is 
shown along with arrows showing the orientation of the magnetic field. The 
effect of the large-scale field on the poloidal loops can be readily seen in 
these simple diagrams. In more quantitative terms,
typical values of $\beta$ in the initial torus lie in the range of 100 to 1000,
for all models: the initial field in the region
where the MRI will operate is weak. The
values of $\beta$ in the region filled with a tenuous radial dust range from
$10^{-3}$ to $10^{-5}$, with the lower values found in the $S$ model.

Each simulation is run for a duration of at least $2000\,M$. This duration is sufficient to pass through the initial transient
phase where initial torus becomes accretion disk (during the first $600\,M$ or so of simulation time), and resolve several complete orbits of the accretion disk, at which time the turbulent transport of angular momentum begins to die down (in axisymmetry). Complete dumps of the code variables are saved every 
$2\,M$ increment in simulation time. Shell-integrated diagnostic values are 
computed and saved every $1\,M$ increment in simulation time.

\begin{table}[htbp]
\caption{\label{params} Torus Simulation Parameters.}
\begin{center}
\begin{tabular}{lll}
\hline
\hline 
\vspace{3pt}
{\bf Black hole:} \\
\quad spin & $a/M$   & 0.90\\
\quad horizon & $r_h/M$   & 1.44\\
\quad marginally stable orbit & $r_{\rm ms}/M$ & 2.32  \\

\hline
\vspace{3pt}
{\bf Torus:} {\footnotesize ({\it See Note a})}\\
\quad inner edge & $r_{\rm in}/M$  & 9.5 \\
\quad pressure maximum& $r_{\rm {P}_{\rm max}}/M$ 
                & 16.1 \\
\quad outer edge & $r_{\rm out}/M$ & 36.0\\
\quad maximum height & $H/M$  & 9.0 \\
\quad orbital period (at $r_{\rm {P}_{\rm max}}$) & $T_{\rm orb}/M$ & 422\\

\hline
\vspace{3pt}
{\bf Dust Background:} {\footnotesize ({\it See Note b})}\\
\quad density contrast & $\bar{\rho}_{\rm dust}/\bar{\rho}_{\rm disk}$
           & $10^{-6}$ \\
\quad energy contrast & $\bar{\epsilon}_{\rm dust}/\bar{\epsilon}_{\rm disk}$
           & $10^{-11}$ \\
               
\hline
\vspace{3pt}
{\bf Grid:} {\footnotesize ({\it See Note c})}\\
\quad Zones & $N_r \times N_\theta$ & $192 \times 192$ \\
\quad Inner boundary & $r_{\rm min}/M$ & 1.45 \\
\quad Outer boundary & $r_{\rm max}/M$ & 120 \\
\quad Polar axis offset & $\Delta \theta/\pi$ & $10^{-3}$ \\
\hline
\vspace{3pt}
{\bf External Field:} {\footnotesize ({\it See Appendix  \ref{initState}})}\\
\quad ``Zero'' field ($Z$) & $B_0$  & $0$  \\
\quad Weak field ($W$) & $B_0$  & $+10^{-8}$  \\
\quad Moderate field ($M_u$) & $B_0$  & $+10^{-2}$  \\
\quad Moderate field ($M_d$) & $B_0$  & $-10^{-2}$  \\
\quad Strong field ($S$)& $B_0$  & $+10^{+2}$  \\
\hline
\hline
\end{tabular}
\end{center}
{\footnotesize
{\bf Note a:} Equations given in Appendix \ref{initState}; $K=0.01$, $q=1.68$, and $\Gamma=$ 4/3.
Torus seeded with poloidal magnetic field loops with 
$\beta=\langle  P_{\rm gas}\rangle/\langle  P_{\rm mag}\rangle \approx 100$. 
Torus also given small (1\%) density perturbation. \\
{\bf Note b:} Equations in Hawley, Smarr \& Wilson (1984);
scaled to grid-averaged density and energy contrast.\\
{\bf Note c:} Radial grid: $\cosh$-scaling, outflow boundaries; $\theta$-grid: exponential scaling, reflecting boundaries.}
\end{table}

\newpage
\section{Summary of Numerical Evolution}

This section outlines the role of the large-scale initial magnetic field in the simulations. 
To better appreciate the interaction between the large-scale field and the initial poloidal 
loops, a brief digression is in order to recapitulate the  
simulations where each type of field is considered separately.

In simulations where the MRI is triggered solely by initial poloidal loops, the
transformation of initial torus into accretion disk proceeds along the lines
discussed extensively by DHK: the initial rapid 
growth of the toroidal magnetic field by shear during the first orbit of the
initial torus; the saturation of the poloidal field MRI,
which progresses outward through the torus; and the evolution by MHD turbulence of the accretion disk. The late-time 
structure can be divided into five regions: 
\begin{itemize}
\item The {\bf main body of the disk}, a turbulent, massive wedge. Gas pressure dominates here and the magnetic and velocity fields
are highly tangled. The outer part of the disk moves radially outward
with time as it gains angular momentum.
\item The {\bf coronal envelope}, a region of low density surrounding
the disk, where gas and magnetic pressure are comparable. 
\item The {\bf inner torus and plunging region}.  The inner torus, lying just outside the marginally stable orbit, temporarily accumulates gas accreting 
from the main disk. The plunging region, where accreting gas spirals into the black hole, lies inward of the inner torus.  
\item The {\bf funnel-wall jet}, an outflow along the centrifugal
barrier which originates near the inner torus.  The density 
here is small compared to the disk, but one to two orders of magnitude 
greater than in the axial funnel. The intensity of the jet is spin-dependent:  
the funnel-wall jet is strongest when black hole spin is near-extremal. The jet is
also fed by entrainment from the adjacent corona.
\item The {\bf axial funnel}, a magnetically-dominated region with tenuous, hot, relativistic outflow. The magnetic field is predominantly radial.
\end{itemize}
Perhaps the most important point in the present context is that the magnetic 
field in these simulations, though initially confined to the torus, eventually
populates the entire simulation volume (Hirose {\it et al.}, 2004);  the 
MRI-driven accretion flow produces its own large-scale field. 
DHK and companion papers documented full 3D simulations; simulations 
carried out in axisymmetry tend to produce a more violent transient phase and the turbulent disk exhibits the channel solution of HB92. However, in both 2.5D and 3D simulations, the above facts remain
essentially unchanged with one important difference: in axisymmetry, the
MRI is not self-sustaining and the rate of accretion dies down within about ten
orbital periods (at the initial pressure maximum).

\newpage
In simulations where a large-scale magnetic field interacts with a radially 
infalling dust, the process which drives the observed unbound outflows is the 
advection and subsequent winding up of the initial magnetic field through frame 
dragging. The time-scale for this process is very short compared to the 
accretion disk simulations: for a Kerr black hole with a spin of $a/M=0.9$, as 
is used in the present simulations, the process of turning the initial infall 
into a broad, unbound outflow takes place within approximately $100\,M$ of 
simulation time. Within this short period, the density of the gas in the 
vicinity of the black hole drops precipitously. The simulation results become 
unreliable when the density and energy approach the numerical floor of the code, 
at approximately $300$ to $400\,M$ in simulation time for the GRMHD code (De 
Villiers, unpublished simulations). As mentioned earlier, the lack of a stable 
mass reservoir near the black hole dooms such simulations to a short life.

\subsection{The Weak-field Case}

\begin{figure}[htbp]
\begin{center}
\includegraphics[width=5.5in]{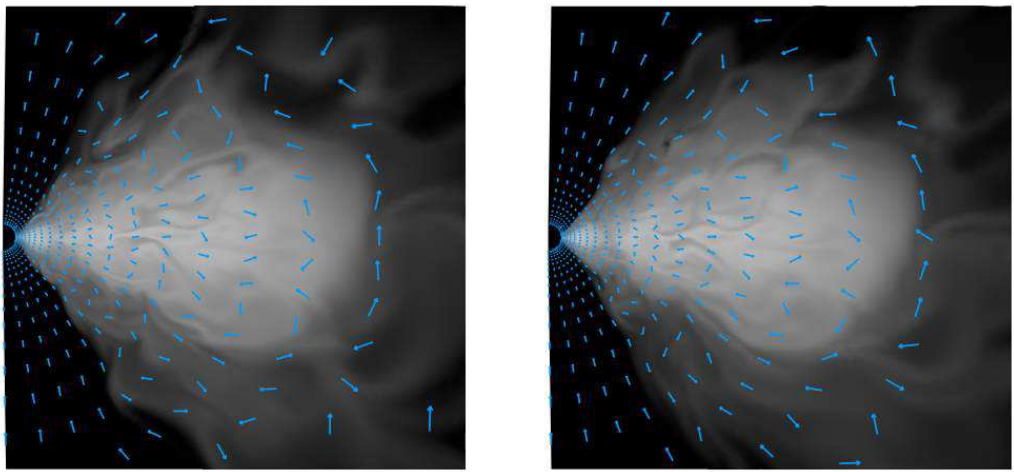}
\caption{{\small Late-time structure ($t/M=1600$) of Weak-field, $W$, simulation (left panel) compared to the ``zero'' field, $Z$, simulation (right panel). Gas density is plotted on a common logarithmic scale. The arrows show the local
direction of the poloidal component of the magnetic field. Each panel spans
a physical extent of $50\,M$; the black hole is centered on the left edge of each panel.}}
\label{WLate}
\end{center}
\end{figure}

The evolution of the $W$ simulation is, in all respects,
very similar to a benchmark simulation with no initial large-scale
field (referred to from hereon as the $Z$, or zero large-scale field 
simulation). Figure \ref{WLate} 
presents the late-time density distribution as well as the orientation 
of the poloidal component of the magnetic field for the $W$ and $Z$ simulations. 
The overall size of the main disk, the shape of the corona, the presence of an
inner torus just outside the marginally stable orbit, and the poloidal magnetic
field structure are virtually identical. As will be seen later, other diagnostics
are also in close agreement for these two simulations. 
\subsection{The Moderate-field Cases}

\begin{figure}[htbp]
\begin{center}
\includegraphics[width=5.in]{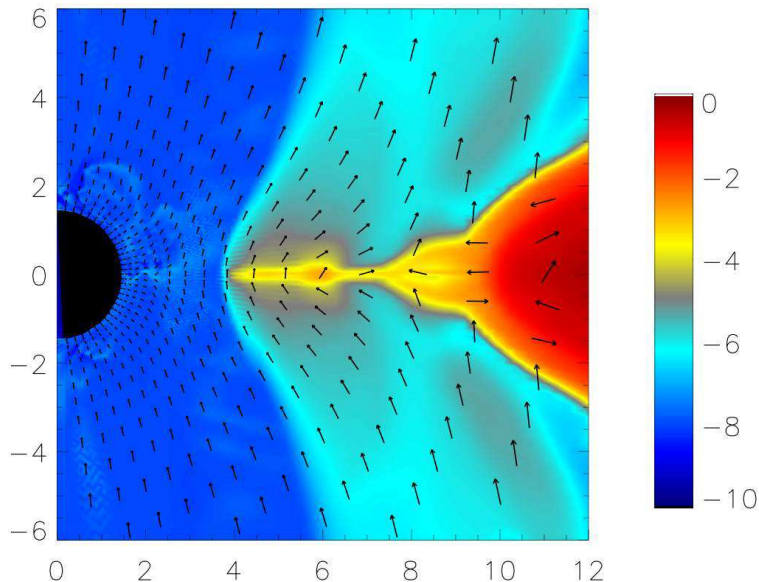}
\caption{{\small Interaction between nascent accretion flow from torus and axial outflow from the wind-up of the initial large-scale field. This image is produced from 
$M_u$ simulation data at $t/M=240$. Gas density is plotted using
a logarithmic colour scale. The arrows show the local
direction of the poloidal component of the magnetic field. The figure spans
a physical extent of $12\,M$; the black hole is centered on the left edge of the panel.}}
\label{MuEarly}
\end{center}
\end{figure}

Differences with the case with no initial large-scale field begin to
emerge in the $M_u$ and $M_d$ simulations. As mentioned above, the time-scale for
the winding up of the large-scale field is very short; in simulations with
only this field and radial dust, an unbound outflow develops promptly 
near the black hole and
impedes further accretion. In the moderate- and strong-field simulations
presented here, this behaviour persists, but coexists with the
early growth phase of the MRI in the torus, which extrudes a thin, dense
current sheet along the equatorial plane. The clash between these two
events is captured in Figure \ref{MuEarly}, which shows the ``shoving match'' between
the nascent accretion flow and the transient outflow from the large-scale field. A sharp boundary between the low-density gas in the prompt axial outflow and the denser material expelled from the initial torus is very obvious. In
all simulations presented here (and also others which probed the parameter
range of the constant $B_0$), the MRI-driven accretion stream always overwhelms
this early outflow. This seems inevitable, if only because the prompt axial outflow, by
its very nature, is a transient effect. This transient is not to be confused with the funnel-wall and axial jets which are {\it long-lived, stable 
features} that arise once the turbulent accretion disk is established.

\begin{figure}[htbp]
\begin{center}
\includegraphics[width=5.5in]{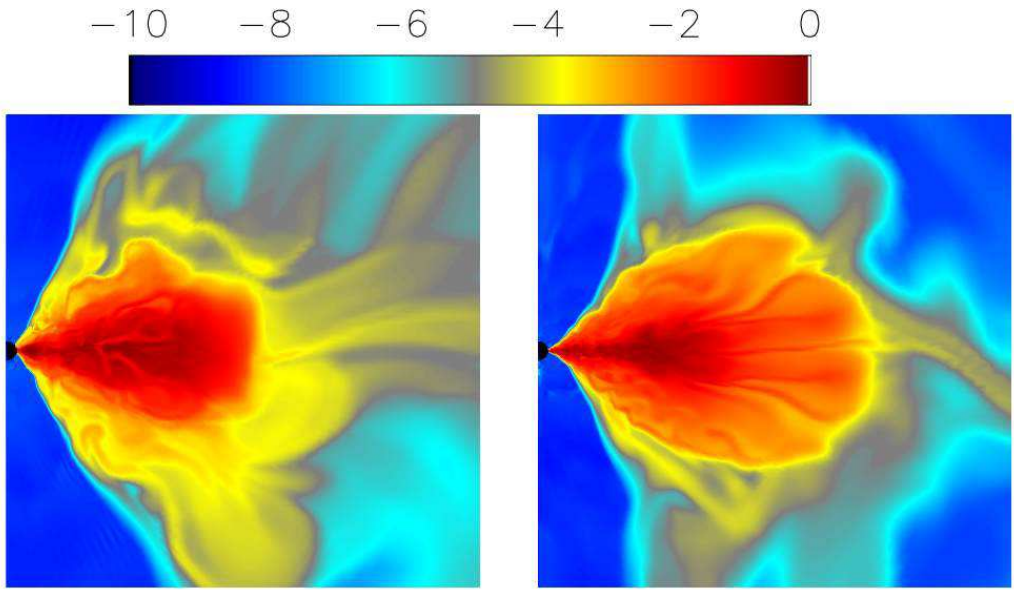}
\caption{{\small Comparison of extensive channels in the Weak-field ($W$; left panel) and Moderate-field ($M_u$) simulations at $t/M=1400$. The presence of extensive
channels at large radii in the main disk is a signature of the simulations with moderate to strong initial large-scale fields. Gas density is plotted using
a logarithmic colour scale. The figures span
a physical extent of $70\,M$; the black hole is centered on the left edge of the panel.}}
\label{Channels}
\end{center}
\end{figure}

Once the accretion flow is established, the overall structure of the disk
resembles previous results: the axial funnel, the funnel-wall jet,
the extensive corona, and the main disk body are generic features of the
simulations, {\it with or without an initial large scale field}. However,
differences do arise in the corona and main disk due to the
presence of the large-scale field and its interaction with the differentially
rotating gas in the main disk body. Since the MRI operates within the initial
torus, which is completely magnetized in these simulations (except for the $Z$
simulation), the MRI has a much more extensive region in which to operate. The
most striking manifestation of this is seen in plots of density, where
extensive channels are observed; see Figure \ref{Channels}. These channels span a much greater 
radial extent in the $M$ and $S$ simulations than in the $W$ simulation.

\subsection{The Strong-field Case}

\begin{figure}[htbp]
\begin{center}
\includegraphics[width=5.5in]{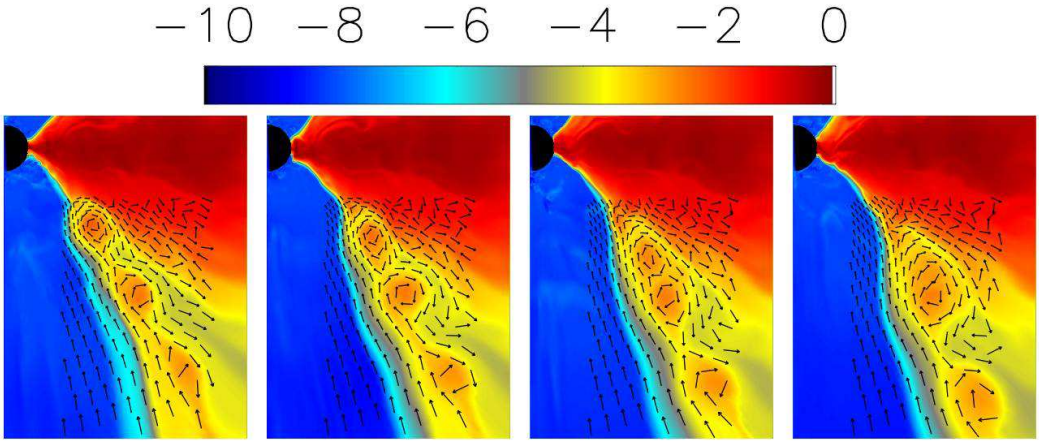}
\caption{{\small Snapshots of density vortices moving away from the black hole following the break-up of a dense channel in the 
$S$ simulation at times (left to right): $t/M=$ 1800, 1808, 1816, and 1824. The 
arrows show the local direction of the poloidal component of the magnetic field 
in and around the vortices; the region surrounding the vortices has not only strong velocity shear (the outflow in the funnel is highly relativistic, whereas gas flow in the corona and funnel wall is much slower), but the 
poloidal component of the magnetic field also features a reversal on either side of the dense stream of gas containing the vortices. Each panel spans
a physical extent of $15\,M$ by $20\,M$; the black hole is located near the upper left corner in each image.}}
\label{Vortices}
\end{center}
\end{figure}

In the $S$ simulation, in which the large-scale field effectively washes out
the initial poloidal loops, the evolution of the initial torus still proceeds along the lines of the other simulations, with similar late-time structures emerging. However, this simulation features much more violent, large-scale  motions of gas in the disk and corona. Animations show extended density channels feeding accreting gas onto the black hole. Those channels lying near
the funnel wall can spontaneously break up into large vortices which move
away from the black hole. Some of these vortices may then fall back and be 
accreted. These vortices appear to evolve, in part, due to the strong velocity 
shear along the funnel wall; they also show a vortical structure in 
the poloidal field, as shown in Figure \ref{Vortices}. Though smaller versions of 
these vortices are observed in all simulations, the $S$ simulation is by far the
most spectacular.

\begin{figure}[htbp]
\begin{center}
\includegraphics[width=5.5in]{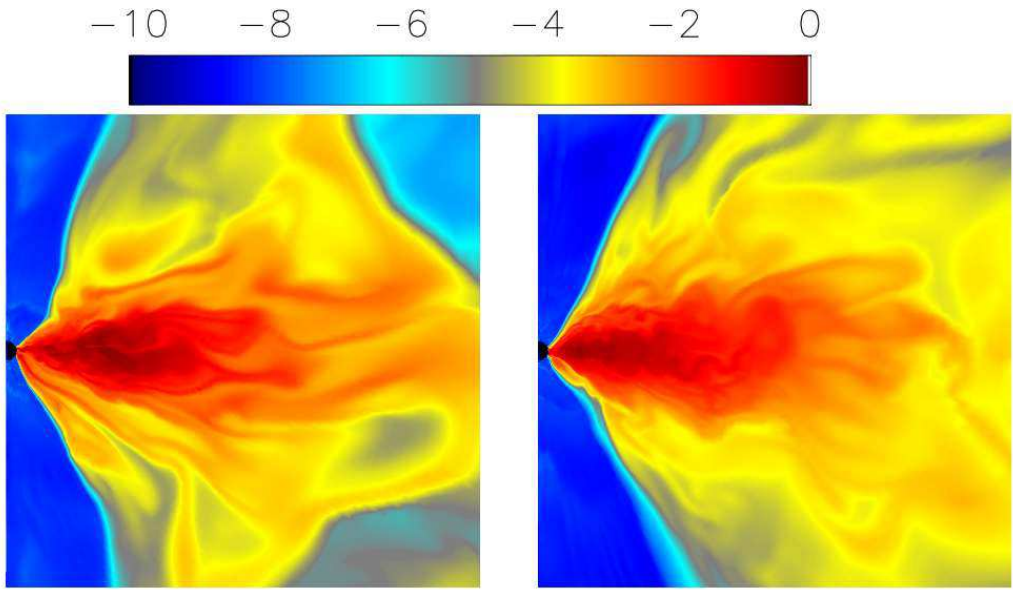}
\caption{{\small Comparison of density at $t/M=2000$ (left panel) to $t/M=4000$ (right panel) in the $S$ simulation. Gas density is plotted using
a logarithmic colour scale. The figures span
a physical extent of $70\,M$; the black hole is centered on the left edge of the panel.}}
\label{RhoS}
\end{center}
\end{figure}

Snapshots of the late-time state of the simulations suggest that the evolution 
of the $S$ simulation is somehow retarded compared to the other simulations; by 
$t/M=2000$, it still exhibits extensive channels in the inner region whereas the 
$W$ and $M$ simulations exhibit a greater degree of turbulence. To investigate 
this, the $S$ simulation was extended to $t/M=4000$. Figure \ref{RhoS} compares 
the density at $t/M=2000$ and $t/M=4000$. It is clear that the main disk 
eventually becomes turbulent, and that a bi-conical funnel and an extensive 
corona also emerge. This reinforces the notion that the outcome of the 
simulations is generic, but also suggests that the presence of the large-scale 
initial field alters the time scale of the evolution somewhat. By the time the
$S$ simulation is fully turbulent, the other simulations, especially the $W$ 
simulation, are showing decreased MRI activity. 

\section{Comparing the Simulations}

In order to compare and contrast the four simulations, records of 
shell-averaged scalars (e.g. density and pressure), as well as shell-averaged 
fluxes (e.g. mass, energy, magnetic stress) are available to
provide a view of the simulations at high time-resolution (these quantities were computed at every $1\,M$ increment in time during the evolution of each disk). In addition,
time-averaging of code variables from the full data dumps provides
a complementary view, by highlighting the more persistent features
of the simulations. 

\subsection{Mass and Energy}

The mass and energy fluxes ($\langle \rho\,U^r\rangle(r,t)$ and 
$\langle {T^r}_t\rangle$(r,t)) quantify the effect of the
large-scale initial field on the accretion rate (at $r_{\rm min}$), and the 
ejection rate of material at large radii. All integrated quantities are computed 
while distinguishing between bound $(-h\,U_t < 1)$ and unbound material; this 
simple distinction helps separate disk material (bound) from
coronal/jet material (unbound).

Table \ref{history} shows the mass and energy passing through the
inner boundary and ejected from the simulation volume through $r/M=100$. 
For reference, the ``Z'' (zero large-scale
field) simulation is also shown to provide a benchmark. 
\begin{table}[htbp]
\caption{\label{history} Mass and Energy Accretion/Ejection.}
\begin{center}
\begin{tabular}{lcccc}
\hline
\hline 
\vspace{3pt}
{\bf Simulation} 
& ${\bf \Delta\,M_{\rm \bf bound}/M_0}$ &${\bf \Delta\,M_{\rm \bf unbound}/M_0}$ 
& ${\bf \Delta\,E_{\rm \bf bound}/E_0}$ &${\bf \Delta\,E_{\rm \bf unbound}/E_0}$\\
\hline
\vspace{3pt}
{\bf Accretion:}\\
$Z$ {\footnotesize (See Note)} 
     & $1.73 \times 10^{-2}$ & $2.84 \times 10^{-4}$ 
     & $1.53 \times 10^{-2}$ & $3.27 \times 10^{-4}$ \\
$W$    & $1.54 \times 10^{-2}$ & $3.34 \times 10^{-4}$ 
     & $1.35 \times 10^{-2}$ & $3.93 \times 10^{-4}$ \\
$M_u$   & $1.44 \times 10^{-2}$ & $2.07 \times 10^{-4}$ 
     & $1.26 \times 10^{-2}$ & $4.58 \times 10^{-4}$ \\
$M_d$   & $2.11 \times 10^{-2}$ & $2.25 \times 10^{-3}$ 
     & $1.78 \times 10^{-2}$ & $2.97 \times 10^{-3}$ \\
$S$    & $2.66 \times 10^{-2}$ & $5.73 \times 10^{-3}$ 
     & $2.25 \times 10^{-2}$ & $7.16 \times 10^{-3}$ \\
\hline
\vspace{3pt}
{\bf Ejection:}\\
$Z$  
     & $7.57 \times 10^{-4}$ & $1.68 \times 10^{-3}$ 
     & $7.73 \times 10^{-4}$ & $2.14 \times 10^{-3}$ \\
$W$    & $6.28 \times 10^{-4}$ & $1.37 \times 10^{-3}$ 
     & $6.39 \times 10^{-4}$ & $1.80 \times 10^{-3}$ \\
$M_u$   & $5.02 \times 10^{-8}$ & $3.31 \times 10^{-4}$ 
     & $2.96 \times 10^{-6}$ & $1.81 \times 10^{-3}$ \\
$M_d$   & $5.19 \times 10^{-8}$ & $4.83 \times 10^{-4}$ 
     & $3.10 \times 10^{-6}$ & $2.03 \times 10^{-3}$ \\
$S$    & $3.94 \times 10^{-8}$ & $8.13 \times 10^{-4}$ 
     & $3.75 \times 10^{-6}$ & $3.58 \times 10^{-3}$ \\
\hline
\hline
\end{tabular}
\end{center}
{\footnotesize
{\bf Note:} This so-called zero-field simulation is a baseline
simulation where the large-scale initial field is switched off,
leaving only the initial poloidal loops with $\beta = 100$.}
\end{table}

The mass/energy accretion and ejection numbers are obtained by
integrating the corresponding shell-averaged fluxes over the
full simulation (details in Appendix \ref{diags}), and normalizing
each to the initial mass/energy of the torus. The accreted
mass and energy is seen to increase with the strength of the
initial large-scale field, though there is some variability in
the models. The rates for models $Z$, $W$, and $M_u$ tend to lie fairly close
to one another. There is a noticeable jump
in the accreted unbound mass and energy for models $M_d$ and $S$, with the
total accreted mass (bound plus unbound) in $S$ increased by a factor of 
$\sim 2$ over the $Z$, $W$, and $M_u$ simulations.

The total amount of energy (bound plus unbound) ejected through $r/M=100$ is slightly enhanced by the initial large-scale field, while ejected mass is lower 
in the $M$ and $S$ simulations than it is in the $Z$ and $W$ cases. The nature 
of the ejected mass/energy, i.e. the portion that is bound vs unbound, shows a
sharp shift in the $M$ and $S$ 
cases, indicating that more of the ejected  material is truly ejected (has 
escape velocity) and would not return into the simulation volume should the 
outer radial boundary be pushed outward.

\begin{figure}[htbp]
\begin{center}
\includegraphics[width=5.5in]{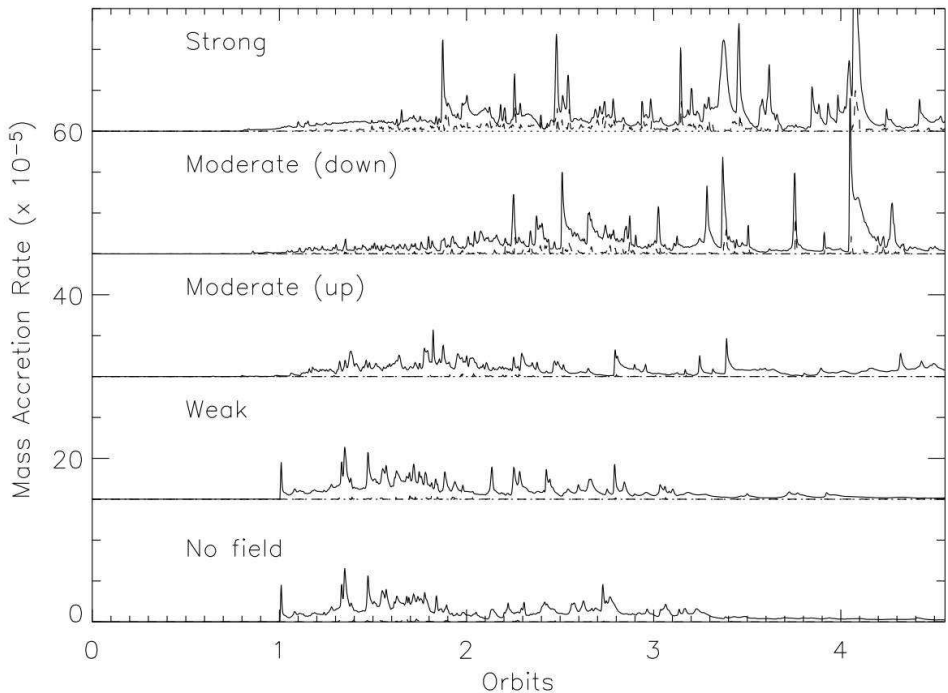}
\caption{{\small Mass accretion rate at the inner radial boundary; each curve is plotted on the same scale, but shifted vertically for clarity (the dashed horizontal 
lines mark the zero for each curve). The solid curves mark the total accretion 
rates (bound plus unbound); while the dashed curves show unbound material only. 
Most of the accreted material is bound.}}
\label{Mdot}
\end{center}
\end{figure}

Mass accretion as a function of time, $\langle \rho\,U^r \rangle(r_{\rm min},t)$, is shown in Figure \ref{Mdot}. This figure 
departs from the convention adopted in this paper and plots simulation time in 
terms of orbital period at the initial pressure maximum. This highlights several 
important features arising from the initial large-scale field:
\begin{itemize}
\item In all simulations, the accretion rate does not pick up until the first orbit is complete: accretion
is driven by the MRI operating within the disk. 
\item The $Z$ and $W$ models show the
characteristic spike at $t=1$ orbit, marking the arrival of the dense accretion
flow at the inner boundary. 
\item The $M$ and $S$ models lack this sharp spike and show a more gradual 
increase, due in part to the ``duel'' between the early, tenuous outflow 
powered by the winding up of the initial large-scale field outside the initial 
torus and the equatorial accretion stream from the torus.
\item The accretion rate of the $Z$ and $W$ models begins to die down after 
three orbits, indicating decreased MRI activity.
\item The accretion rate in the $M_u$ model shows some diminution after the 
third orbit, but remains stronger than the $W$ and $Z$ models.
\item The accretion rate in the $M_d$ and $S$ models shows no sign of decreasing 
after the third orbit, and in fact seems to be growing in strength; the large 
spikes correspond to the arrival of dense channels and/or inbound vortices, 
similar to those shown in Figure~\ref{Vortices}.
\end{itemize}

\newpage
\subsection{Time-averaged Data}

The overall structure of the main disk, corona, and funnel is best
viewed in time-averaged plots, which smooth out the often intense variability
of the simulations. 
The three main differences between the simulations are seen in the density 
distribution in the corona and axial funnel, in the 
overall pressure distribution, and in the relative strength of the toroidal
and poloidal components of the magnetic field.

\begin{figure}[htbp]
\begin{center}
\includegraphics[width=5.5in]{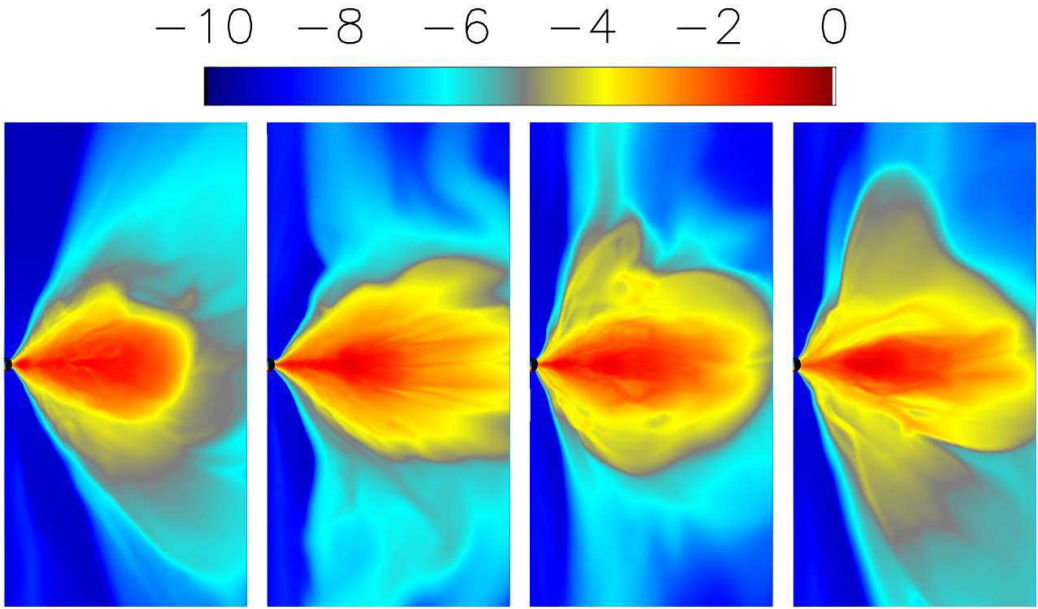}
\caption{{\small Comparison of time-averaged (from $t/M=$1200 to 1600) density for all simulations, from left to right: $W$, $M_u$, $M_d$, $S$. Logarithmic colour scale is normalized to initial maximum density. Figures span
a width $50\,M$ and a height of $100\,M$; the black hole is centered on the left edge of the panel.}}
\label{rhobar}
\end{center}
\end{figure}

Figure \ref{rhobar} shows the density distribution in each simulation, averaged 
from $t/M=$ 1200 to 1600. This figure shows the more 
expansive region occupied by the main disk in the $M$ and $S$ simulations, due
to the action of the MRI on the more extensive initial magnetic field. The 
figure also shows that the $M$ and $S$ simulations tend to produce a narrower 
funnel, and hence a more collimated axial jet; the figure suggests that a tight, 
nearly cylindrical outflow is achieved. However, Figure \ref{RhoS} suggests that 
this is a transient phenomenon: the magnetic structure at large radii appears to 
evolve over a longer time in the $M$ and $S$ simulations, and the tight 
collimation eventually vanishes.

\begin{figure}[htbp]
\begin{center}
\includegraphics[width=5.5in]{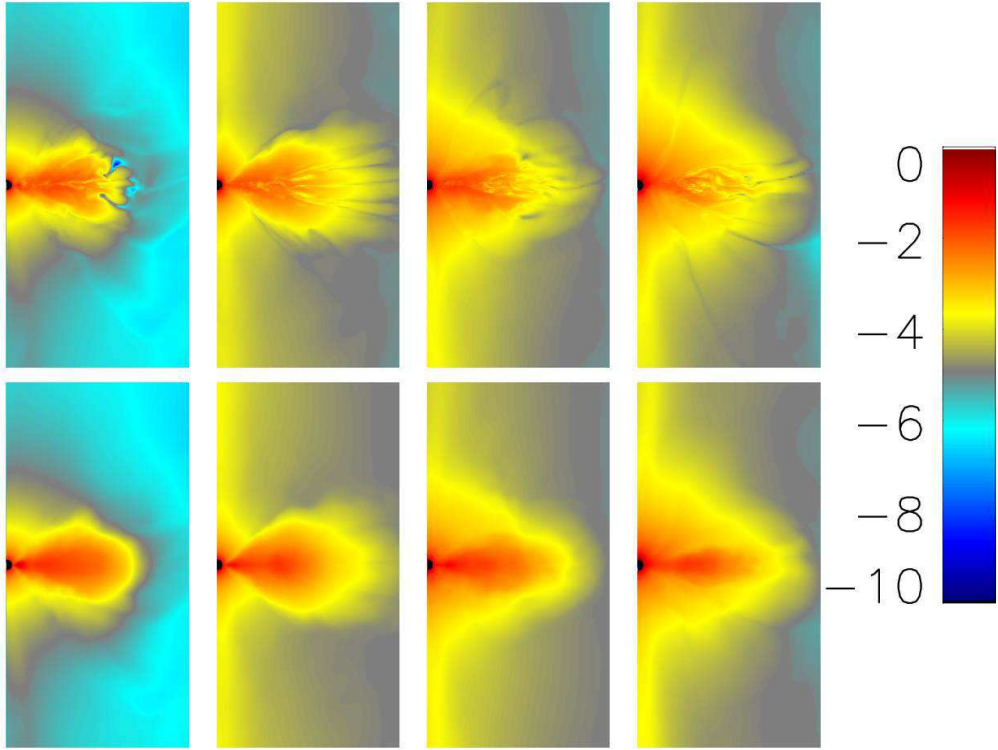}
\caption{{\small Comparison of time-averaged pressure (from $t/M=$1200 to 1600) for all simulations, from left to right: $W$, $M_u$, $M_d$, $S$. Top row shows magnetic pressure; bottom row shows
total pressure (gas and magnetic). The logarithmic colour scale is normalized to maximum total pressure at $t/M=1600$ for simulation S. The figures span
a width $50\,M$ and a height of $100\,M$; the black hole is centered on the left edge of the panel.}}
\label{Pbar}
\end{center}
\end{figure}

\newpage
Figure \ref{Pbar} shows the pressure distribution in each simulation,
averaged from $t/M=$ 1200 to 1600.
Both the total pressure (gas plus magnetic) as well as the magnetic pressure
are shown. The $W$ simulation has a total pressure distribution essentially
identical to simulations with no initial large-scale magnetic field, namely a
spherical ``halo'' structure in the funnel and a horizontally stratified 
structure in the main disk. The $M$ and $S$ simulations show a marked departure
from this, with a significantly stronger magnetic pressure in the corona and
at large radii in the funnel. The region of high magnetic pressure in 
the coronal region immediately above and below the inner torus in the $M_d$ and 
$S$ simulations has a scale height greater than that of the main disk. 
Evidence that the $M$ and $S$ simulations are
less evolved at large radii is also found in these plots: the pressure distribution at large radii still retains some of its initial structure.
A region of high magnetic pressure along the spin axis
is likely an artifact of the reflecting boundary conditions used in axisymmetry. 

\subsection{Evolution of the Magnetic Field}

\begin{figure}[htbp]
\begin{center}
\includegraphics[width=5.5in]{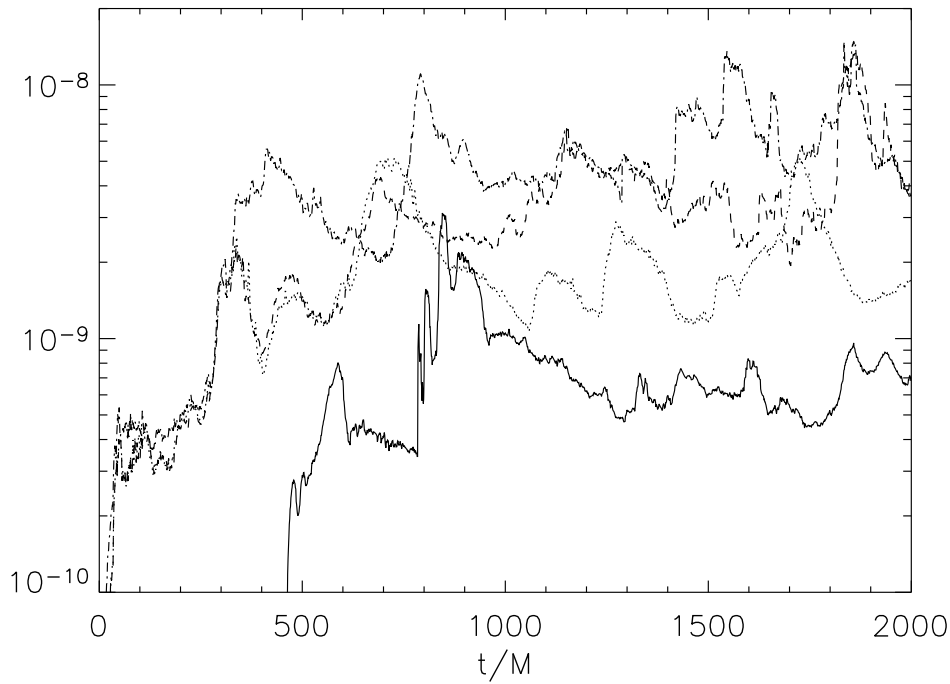}
\caption{{\small Time-evolution of the shell-averaged, unbound magnetic pressure at $r/M = 16.1$. Unbound material is found in the corona and the axial jets, and
the radial distance of the shell shown here passes through the initial pressure maximum. The solid line denotes the $W$ simulation; the dotted line denotes $M_u$; the dashed line, $M_d$; the dash-dotted line, $S$.}}
\label{MagEvolution}
\end{center}
\end{figure}

Since the material in the corona and the axial jets is entirely unbound 
$(-h\,U_t > 1)$, the diagnostic tracking the evolution of magnetic pressure in 
unbound gas provides a view of the time-evolution of the magnetic field in these 
regions. Figure \ref{MagEvolution} shows the time-evolution, for all four simulations,
of unbound magnetic pressure on the shell passing through the initial pressure 
maximum, a region where Figure \ref{Pbar} indicates a pronounced difference 
between simulations. The $M$ and $S$ simulations show a prompt rise in unbound 
magnetic pressure, due to the stronger large-scale initial field, which is 
wound up by frame-dragging. The growth of unbound magnetic pressure in the $W$ 
simulation lags considerably, reinforcing the idea that this simulation is driven by the MRI acting on the initial poloidal loops, that is that unbound 
material is produced during the saturation phase of the MRI, as it is in 
``conventional'' disk simulations. Furthermore, after the initial sharp rise, 
the magnetic pressure for the $M$ and $S$ simulations continues to increase 
steadily with time, whereas the magnetic pressure in the $W$ simulation reaches 
a peak at $t/M \approx 850$, and then decreases. 

\begin{figure}[htbp]
\begin{center}
\includegraphics[width=5.5in]{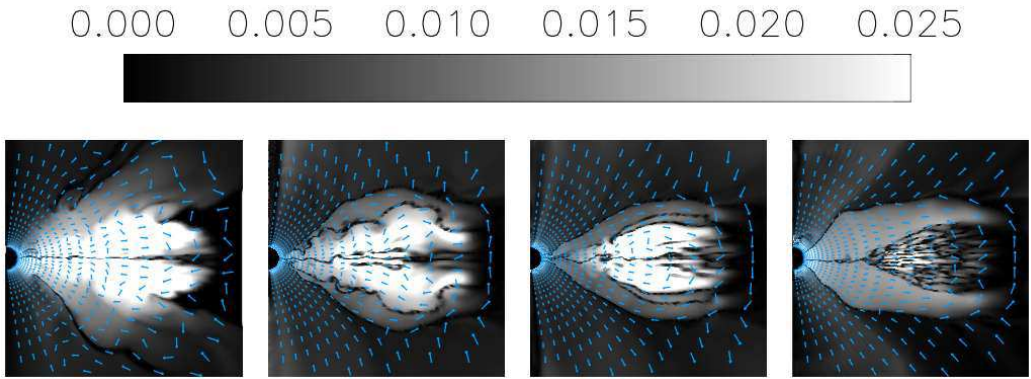}
\caption{{\small Snapshot of magnetic field at $t/M = 330$ for all simulations, from left to right: $W$, $M_u$, $M_d$, $S$. The arrows show the local
direction of the poloidal component of the magnetic field; the toroidal
component is shown in grey-scale, in code units. Each panel spans
a physical extent of $30\,M$; the black hole is centered on the left edge of each panel. }}
\label{TorField}
\end{center}
\end{figure}

In the $M$ and $S$ simulations, there is a sharp jump in magnetic pressure at
$t/M \sim 300$, with the greatest jump occurring for the $S$ simulation. A 
snapshot of the magnetic field at this early time is given in Figure \ref{TorField}.
This figure shows the toroidal component of the magnetic field plotted in grey 
scale, along with arrows showing the local direction of the poloidal field. 
The $S$ simulation stands out in this figure: the region between the thin
equatorial inflow and the funnel wall is occupied by a region of intense,
smooth toroidal field. In addition, while the $W$ and $M$ models show regions of 
strong radial streams of toroidal field near the initial pressure maximum, 
indicating the growth phase of the MRI, the $S$ simulation shows exactly the 
opposite, with a region of weaker radial streams at much greater radii. The 
sequence of images suggests that the stronger initial large-scale field 
generates a sheath of toroidal field around the torus which squeezes the region 
of MRI activity to larger radii. This reinforces the notion that 
the evolution of the $S$ simulation is in many respects retarded compared to the 
other simulations. 
Another important feature in this figure is the presence of toroidal field in 
the axial funnel, with the strongest funnel field found in the $S$ simulation. 
Animations show that this toroidal component of the funnel field is highly 
time-variable, appearing in bursts that suggest the launching of torsional waves 
through the funnel. These bursts originate near the inner radial boundary, 
indicating an origin in the ergosphere.

\begin{figure}[htbp]
\begin{center}
\includegraphics[width=5.5in]{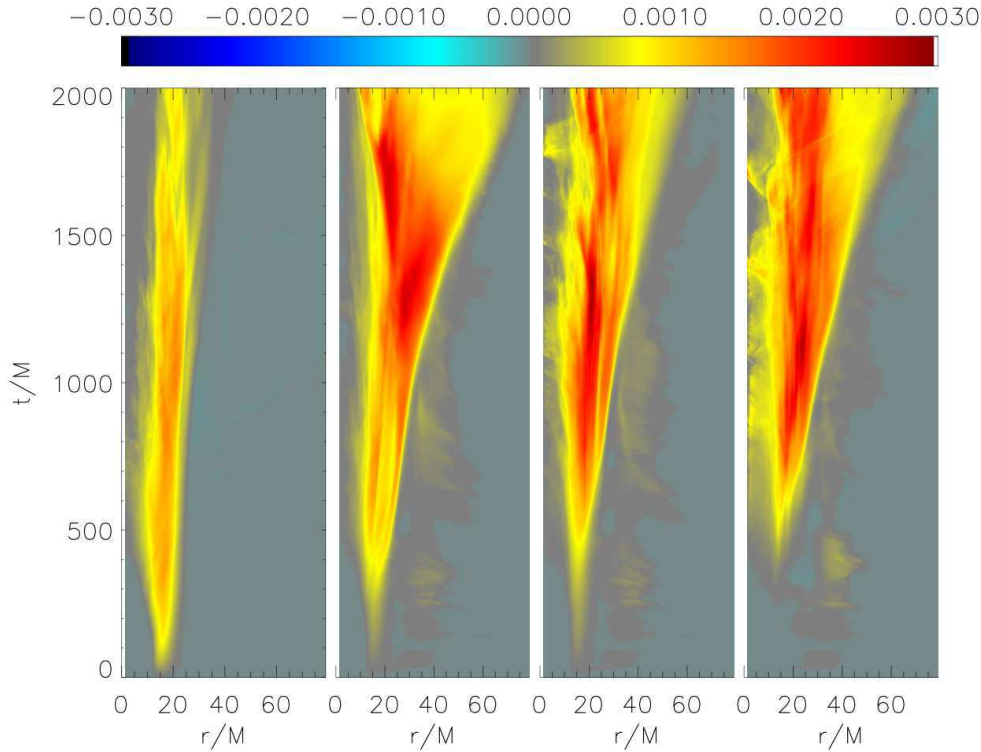}
\caption{{\small Spacetime diagrams of magnetic stress (bound material), 
$\langle \left(b^r\,b_\phi\right)|_{-h\,U_t >1}\rangle(r,t)$, from left to right: simulations
$W$, $M_u$, $M_d$, $S$.}}
\label{stress}
\end{center}
\end{figure}

Finally, to discuss the onset of turbulence in the main disk, take
the $b^r\,b_\phi$ component of the shell-averaged angular momentum flux
(bound component), $\langle \left(b^r\,b_\phi\right)|_{-h\,U_t >1}\rangle(r,t)$. Figure \ref{stress} shows spacetime diagrams for this quantity for each of the 
four simulations. The $W$ model shows a steady growth in the first few $100\,M$ 
of simulation time, followed by a quasi-steady state, then a decrease after 
$1500\,M$. The region of elevated magnetic stress is confined to the vicinity of 
the initial pressure maximum, where the initial magnetic field is concentrated. 
The decrease in stress at late times is consistent with the decrease in 
accretion rate seen in Figure \ref{Mdot}, suggesting a decline in  MRI activity. 
The $M$ and $S$ simulations show two important differences from the $W$ 
simulation: the onset of elevated stress is delayed, with the delay most 
pronounced in the $S$ simulation; and the region of elevated stress expands with 
time to large radii, engulfing a greater portion of the disk. The region of elevated stress extends to the end of the 
simulation in the $M_d$ and $S$ simulations, suggesting that the turbulent phase 
of the MRI is much longer-lived when a strong initial large-scale field is 
present.

\begin{figure}[htbp]
\begin{center}
\includegraphics[width=1.5in]{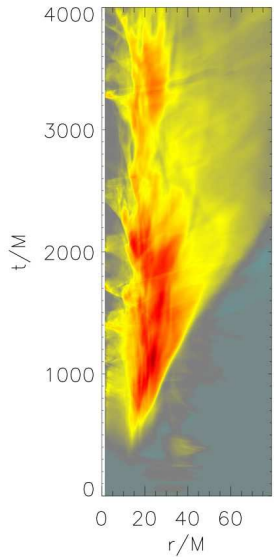}
\caption{{\small Spacetime diagram of magnetic stress (bound material), 
$\langle \left(b^r\,b_\phi\right)|_{-h\,U_t >1}\rangle(r,t)$,  for the extended $S$ simulation.}}
\label{stress2}
\end{center}
\end{figure}

The duration of the growth phase for stress in the $S$ simulation can
investigated by considering the records from the extended $S$ simulation.
Figure \ref{stress2} shows the extended diagnostic for this simulation: the
period of growing/elevated stress, though considerably longer in the $S$
simulation, is nonetheless finite in duration. This simulation shows a
decrease in intensity after $t/M \sim 2300$. 

\section{Discussion\label{discuss}}

This paper has presented results of a group of axisymmetric
GRMHD simulations of accretion disks embedded in a large-scale
magnetic field. To investigate the role of this large-scale
field, all simulations used a common initial state: Kerr black hole
with spin $a/M=0.9$, equilibrium 
torus perturbed by small, random density fluctuations, and an MRI 
seed field consisting of weak poloidal loops confined to the core 
of the torus. The simulations differed only in the scaling constant 
for the initial large-scale field. This initial field was termed 
weak, moderate, or strong, in relation to its effect on the
loops of poloidal field: weak if it left the loops intact;
moderate if it cancelled the portion of the loops anti-parallel to it, on 
either side of the initial pressure maximum; and strong, if it
overwhelmed the loops.

The large-scale initial magnetic field does not alter the general 
features of the late-time structure of the accretion system: the features
identified in DHK (main disk, inner torus, plunging region, coronal
and axial funnel) remain qualitatively the same.
However, the relative strength of the initial large-scale field was found
to have an effect on the details of the evolution of the initial torus. The key
points are the following:
\begin{itemize}
\item the stronger the field, the greater the accretion rate, though the
effect is modest;
\item the stronger the field, the greater the ejection rate of energy,
though the effect is, again, modest;
\item the stronger the field, the greater the fraction of ejected material
that is unbound;
\item the stronger the field, the greater the scale height of the region of elevated 
magnetic pressure in the corona;
\item the stronger the field, the later the onset of the growth phase
of the MRI;
\item the stronger the field, the more intense the density channels prior to the
onset of turbulence.
\end{itemize}
In addition, the large-scale field, due to the higher coronal magnetic pressure, 
tends to produce a more collimated jet, at least in the early stages, 
effectively by squeezing low-density funnel gas closer to the spin axis (Figure 
\ref{TorField}). This suggests that a 
cylindrically-collimated jet is produced very close to the black hole, 
though simulations with a
larger radial grid should be carried out to verify whether this
collimation is found at large distances. Indications here are that the
collimation is confined to the region near the black hole, and that it
is short-lived; a more conical shape is achieved at late times in the
extended $S$ simulation.

The mass and energy accretion and ejection rates in Table \ref{history} can 
be compared to those
found in earlier simulations, notably those in DHK and D05a. It is possible to compare the $Z$ simulation to its closest D05a counterpart, the 3D model KDPi 
(KDP with $\Gamma=4/3$). The rates are found to be systematically 
lower in the simulations discussed here, by a factor of $\sim 2$. The most important difference which can account for the lower rates is the use of axisymmetry, which tends to reduce the accretion rate (De Villiers \& Hawley, 2003; DH03b). The rates given in Table \ref{history} can also be
compared to the 3D simulation in D05b, model $R_{\rm vf\,3D}$, which is very 
similar to the $W$ simulation discussed here. The mass accreted onto the black 
hole for this earlier simulation is lower than that found here, by a factor of 
$\sim 5$, whereas the ejected quantities are systematically higher 
than those shown here, again by a factor of $\sim 5$.

Hawley \& Balbus (1992) discussed the evolution of a shearing sheet
containing an initially weak vertical magnetic field, and
documented the emergence of the channel solution, radially 
elongated regions of elevated density sandwiched between regions of low
density and strong magnetic field, as a direct consequence of the MRI
interacting with the vertical field. In subsequent studies of
axisymmetric disks, these channels have emerged as a 
characteristic feature (channels also arise in 3D, but quickly dissipate;
DH03b). Channels are found in the set of simulations discussed here, and the 
channels are enhanced in strength and radial extent by the presence of the
pervasive large-scale field. This is not surprising: the large-scale
initial field permeates the entire initial torus, and the MRI is thus
provided with a much larger initial region in which to amplify
magnetic field. The simulations with a stronger initial large-scale field give 
rise to a more vigorous evolution, a feature better appreciated by viewing 
animations of gas density, though the figures presented here capture some of 
this detail.

What may seem paradoxical is that the strongest initial field
simulated here shows the latest onset of turbulence: though the
$S$ simulation is by far the most intense, the growth and subsequent
saturation of the MRI are the slowest of all the simulations. However, 
this is not a new finding: HB92 (c.f. Fig. 5) observed the same 
effect in the shearing sheet simulations, namely that stronger initial fields 
show a delayed onset of the growth phase of the MRI when all other
simulation parameters are kept constant. HB92 also
found that stronger initial magnetic fields generate a much stronger maximum 
field, {\it provided that the initial field was not so strong as to
be stable against the MRI}. The current simulations are consistent with these 
earlier findings, for the same reasons: though dealing with a more
complicated initial state, the $Z$, $W$, $M$, and $S$ simulations differ only in 
the strength of the initial large-scale field, and all other parameters (grid 
size, time step size, disk configuration) are identical. In addition, the
strength of the initial large-scale field was kept to a reasonable
level, that is with $\beta \gg 1$ in the initial torus, in recognition of the 
fact that the importance of the MRI as the mechanism of angular momentum
transport in astrophysical disks rests on the fact that it operates on 
{\it weak} magnetic fields. Simulations that begin with inordinately
strong magnetic fields may well preclude the development of the MRI and
may, as was observed in some early tests for the present study, severely disrupt
the initial torus to the point where the physical plausibility of the simulation 
is in doubt.

Another interesting finding is the noticeable difference in outcomes
between the $M_d$ and $M_u$ simulations which differ only in the relative
orientation of the large-scale initial field. The outcome of the $M_u$
simulation, where the field is in the usual ``up'' sense, is very 
similar to the weak-field $W$ simulation, whereas the $M_d$ simulation,
where the field is oriented ``downward'', bears a closer similarity to the
strong-field $S$ simulation. The different outcomes seem to be rooted in
the interaction between the large-scale initial field and the poloidal
loops: the $M_u$ simulation disrupts the poloidal loops in such
a way as to strengthen the field in the narrow region near the inner edge of 
the torus, but it also significantly weakens the field at the center of the 
torus (just outside the initial pressure maximum). In this sense, the $M_u$
simulation begins from a weaker net initial field. The $M_d$ simulation
does exactly the opposite, by reinforcing the poloidal loops at the
center of the torus it produces a relatively stronger initial
field.  

There are many aspects to these simulations that were only briefly
touched upon in this survey. More detailed analysis of the simulation
data should be undertaken, as well as complementary simulations to probe
the effect of other simulation parameters, most importantly black hole spin. 
Further investigation is warranted in the properties
of the funnel: the transient collimation near the base of the funnel 
due to enhanced coronal pressure; the appearance of bursts of toroidal
field in the funnel and their presumed association to torsional
Alfven waves; and the behaviour of the axial jets at large radii. 
More importantly, the lifting of the axisymmetry constraint by proceeding
to full 3D simulations is essential to answer questions regarding the
long-term properties of the accretion disks when immersed in large-scale
fields.

\newpage 
\noindent{\bf Acknowledgements:}

\noindent I thank John Hawley for ongoing discussions on the simulations. 
This research was carried out while at the University of Calgary, Department
of Physics and Astronomy, and was
supported by the Natural Sciences and Engineering Research
Council of Canada (NSERC), as well as the Alberta Ingenuity Fund (AIF).
The simulations were carried out on the IBM system {\it cortex}, a
WestGrid facility. WestGrid computing resources are funded in part by the 
Canada Foundation for Innovation, Alberta Innovation and Science, BC Advanced Education, and the participating research institutions.

\appendix

\section{Equations of GRMHD\label{GRMHDequations}}

The GRMHD code, described in its original form in DH03a, uses a finite difference approach to solve the equations of 
ideal general relativistic magneto-hydrodynamics in the 
spacetime of a Kerr (rotating) black hole in Boyer-Lindquist 
coordinates, $(t,r,\theta,\phi)$. 
The GRMHD code  
solves the equation of continuity, $\nabla_\mu \left(\rho\,U^\mu\right)=0$,
the energy-momentum conservation law, $\nabla_\mu\,T^{\mu \nu} = 0$,
and the Maxwell's equations, 
$\nabla_\mu F^{\mu \nu} = 4\,\pi\,J^\nu$ and 
$\nabla_\mu {}^*F^{\mu \nu} = 0$, which, in the MHD approximation,
reduce to the induction equation
$\partial_\delta\,F_{\alpha \beta} + 
\partial_\alpha\,F_{\beta \delta} + 
\partial_\beta\,F_{\delta \alpha} = 0$. 
In the above expressions,
$\rho$ is the density, $U^\mu$ the 4-velocity, 
$T^{\mu \nu}$ the energy-momentum tensor, and
$F_{\alpha \beta}$ the electromagnetic field strength tensor.
The energy momentum tensor is given by 
$T^{\mu \nu} = \left[ \left(\rho\,h+{\|b\|}^2\right)\,U^\mu\,U^\nu + 
\left(P+{{\|b\|}^2 \over 2}\right)\,g^{\mu \nu} - b^\mu\,b^\nu\right]$
where $h=1 + \epsilon + P/\rho$ is the specific 
enthalpy, with $\epsilon$ the specific internal energy and 
$P=\rho\,\epsilon\,(\Gamma-1)$ the ideal gas pressure ($\Gamma$ is the
adiabatic exponent); ${\|b\|}^2=b^\mu\,b_\mu$ is the magnetic field 
intensity; and $b^\mu = {}^*F^{\mu \nu}\,U_\nu/(4\,\pi)$ is the
magnetic field 4-vector. 
The induction equation is rewritten in terms of the Constrained Transport
(CT; Evans \& Hawley, 1988) magnetic field variables 
${\cal B}^i = \epsilon_{i j k}\,F_{j k}$,
as $\partial_t {\cal B}^i - 
\partial_j\left( V^i\,{\cal B}^j - V^j\,{\cal B}^i\right)= 0$,
where $V^i=U^i/U^t$ is the transport velocity, and $U^t=W/\alpha$,
with $W$ the Lorentz factor. The CT algorithm ensures that the
constraint $\partial_i {\cal B}^i = 0$ is satisfied to rounding
error. 
The Kerr metric is expressed in Boyer-Lindquist coordinates, for
which ${ds}^2=g_{t t}\,{dt}^2+2\,g_{t \phi}\,{dt}\,{d \phi}+g_{r r}\,{dr}^2 +
g_{\theta \theta}\,{d \theta}^2 +g_{\phi \phi}\,{d \phi}^2$;
$\alpha = {(-g^{tt})}^{-1/2}$ is the lapse function.
Geometrodynamic units are used, where $G = c = 1$, and time 
and distance are given in units of the black hole mass, $M$.  

A recent effort at development and testing of the GRMHD code has yielded
improved stability in highly magnetized flows by adding artificial
diffusion terms to the equations of continuity, energy conservation, and 
momentum conservation. Though a detailed discussion of these tests is beyond
the scope of this paper, the reason for adding such diffusion terms
is quite simple: truncation error due to finite differencing of time derivatives 
in expressions of the form
\begin{equation}\label{eqA}
\partial_t{{\cal X}} + {1 \over \sqrt{\gamma}}\partial_j
\left( \sqrt{\gamma}\,V^j\,{\cal X}\right)
\end{equation}
gives rise to grid noise that may be amplified to unacceptable levels
over time. The remedy is to add compensatory flux terms which cancel the
leading term in the remainder of a formal Taylor expansion of the
time derivative operator.
The time derivative term, expanded to second order, is given here in
schematic form:
\begin{equation}
{\left(\partial_t{{\cal X}}\right)}_i \approx 
{ {\cal X}^{n+1}_i - {\cal X}^{n}_i \over \Delta t} - 
{\Delta t \over 2}{\left(\partial^2_t{{\cal X}}\right)}_i\, ,
\end{equation}
where the superscripts denote time levels and the subscript denotes
a spatial grid index. Using (\ref{eqA}), the second-order term can be
rewritten as
\begin{equation}
\partial^2_t{{\cal X}} \approx 
{1 \over \sqrt{\gamma}}\partial_j \left[V^j
\left( \partial_i\sqrt{\gamma}\,V^i\,{\cal X}\right)\right]
\end{equation}
and fed back in to the discretized form as an artificial diffusion
term:
\begin{equation}
{ {\cal X}^{n+1}_i - {\cal X}^{n}_i \over \Delta t} \approx 
-{1 \over \sqrt{\gamma}}\partial_j \left[\sqrt{\gamma}\,V^j\left\{
{\cal X}^{n}_i-
{a\,\Delta t \over 2\,\sqrt{\gamma}} \left( \partial_k\sqrt{\gamma}\,V^k\,
{\cal X}^{n}_i\right)\right\}\right].
\end{equation}
Numerical tests have shown that the diffusion constant $a = 6 \times 10^{-3}$ yields good results.

These modifications were important for the simulations discussed here since
the addition of the initial large-scale magnetic field produces conditions 
near the black hole where shocks in a highly magnetized flow can give rise to numerical artefacts.  However, it was also found
that these modifications to the GRMHD code had only a minor impact on ``conventional'' accretion disk simulations (i.e. those without an initial large-scale magnetic field) such as the KD models discussed in DHK and
companion papers. This
is corroborated by the agreement mentioned in \S \ref{discuss} between
the present simulations and those in D05a. The
version of the GRMHD code used to produce DH05b, an intermediate form
of the code, tended to overestimate funnel energetics.

\section{\bf Details on the Initial State\label{initState}}

The initial torus, discussed in detail in DHK, is an equilibrium disk solution 
to the equations of GR hydrodynamics which is then pertubed with a small, random 
density fluctuation and overlaid with a magnetic field that serves as a seed 
field for the MRI.

The equilibrium solution is found by considering a disk with a power-law 
rotation,
\begin{equation}\label{kd.1}
\Omega = \eta\,\lambda^{-q}
\end{equation}
where 
\begin{equation}\label{kd.7}
\lambda^2 = {l \over \Omega} 
 = l {\left(g^{t\,t}-l\,g^{t\,\phi} \right) \over 
      \left(g^{t\,\phi}-l\,g^{\phi\,\phi} \right)}\, ,
\end{equation}
$\eta$ is a constant, and $q$ a 
positive number. The hydrodynamic momentum equation is (Hawley, Smarr, \& 
Wilson, 1984)
\begin{equation}\label{kd.2}
\partial_t\left(S_j\right)+
  {1 \over \sqrt{\gamma}}\,
  \partial_i\,\sqrt{\gamma}\,\left(S_j\,V^i\right)+
  {1 \over 2}\,\left({S_\epsilon\,S_\mu \over S^t}\right)\,
  \partial_j\,g^{\mu\,\epsilon}+
  \alpha\,\partial_j\left(P\right) = 0 ,
\end{equation}
where $S_j = \rho\,h\,W\,U_j$ is the momentum.  

The momentum equation is simplified by imposing time-independence, 
axisymmetry, and requiring that there be no poloidal motion.  This 
simplified equation reads,
\begin{equation}\label{kd.3}
  \alpha\,\partial_j\left(P\right)+
  {1 \over 2}\,\left({S_\epsilon\,S_\mu \over S^t}\right)\,
  \partial_j\,g^{\mu\,\epsilon} = 0 .
\end{equation}
Use the definition of the momentum $4$-vector and the definition
of specific angular momentum, $l =-U_{\phi}/U_t$, to obtain
\begin{equation}\label{kd.4}
  {\partial_j\left(P\right) \over \rho\,h} = 
 -{{U_t}^2 \over 2}\,\partial_j\left({U_t}^{-2}\right)+
  {U_t}^2\,\left(-\partial_j\,g^{t\,\phi}+l\,\partial_j\,g^{t\,\phi} \right)
  \,\partial_j\,l,
\end{equation}
where ${U_t}^{-2}=g^{t\,t}-2\,l\,g^{t\,\phi}+l^2\,g^{\phi\,\phi}$.

Assume constant entropy, $T ds = 0$, so that $dh=dp/\rho$.
Use (\ref{kd.1}) and (\ref{kd.7}) to write 
$l = \Omega \lambda^2 = c\,\lambda^{2-q}$ and 
$\Omega = c^{-2/(q-2)}\,l^{q/(q-2)} \equiv k\,l^\alpha$ . So (\ref{kd.4}) can be written in integral form
\begin{equation}\label{kd.6a}
  {\int}_{h_{in}}^{h} {dh \over h}= 
 -{1 \over 2}\,{\int}_{U_{in}}^{U_t} {d\left({U_t}^{-2}\right)\over {U_t}^{-2}}+
  {\int}_{l_{in}}^{l} {k\,l^\alpha \over 1- k\,l^{\alpha+1}}\,d\,l
\end{equation}
where $\alpha = q/(q-2)$ and 
$x_{in}$ refers to the quantity in question evaluated on the surface
of the disk. Clearly, $h_{in}=0$, and a general solution admitting a
choice of surface binding energy $U_{in}$ is given by
\begin{equation}\label{kd.10}
  h(r,\theta) = { U_{in} f(l_{in}) \over U_{t}(r,\theta) f(l(r,\theta))} ,
\end{equation}
where $f(l) = {\|1 - k\,l^{\alpha+1}\|}^{1/(\alpha+1)}$ or
$f(\Omega) ={\|1 - k^{-1/\alpha}\,\Omega^{(\alpha+1)/\alpha}\|}^{1/(\alpha+1)}$.
Using the equation of state and the definition of enthalpy, the
internal energy of the disk is
\begin{equation}\label{kd.11}
  \epsilon(r,\theta) = {1 \over \Gamma} \left(
  { U_{in} f(l_{in}) \over U_{t}(r,\theta) f(l(r,\theta))}-1\right) .
\end{equation}
For a constant entropy adiabatic gas the pressure is given by $P =
\rho\,\epsilon\,(\Gamma - 1) = K\,\rho^\Gamma$, and density is given by
$\rho={\left[{\epsilon\,(\Gamma - 1) / K}\right]}^{1/(\Gamma - 1)}$.

For the Schwarzschild metric, these analytic relations completely specify the
initial equilibrium torus.  In the Kerr metric, $\lambda^2$ is an implicit
function of $l$, requiring an iterative approach to solve for the equilibrium state.
However, by using the Schwarzschild expression for $\lambda^2$ in
the Kerr metric, the above results can be used to initialize a disk that
maintains a sufficiently good equilibrium to use an initial state for numerical
studies of the MRI.

Poloidal loops of magnetic field are laid
down along isodensity surfaces within the torus by defining
$A_{\mu} = (A_t,0,0,A_\phi)$, where
\begin{equation}\label{vecpot}
A_{\phi\,{\rm (loops)}} = 
\cases{
k (\rho-\rho_{cut}) & for $\rho \ge \rho_{cut}$ \cr
0 & for $\rho < \rho_{cut}$},
\end{equation}
where $\rho_{cut}$ is a cutoff density corresponding to a particular isodensity
surface within the torus.  The
constant $k$ is set by the input parameter $\beta$, the ratio of the gas
pressure to the magnetic pressure.
The constant $\rho_{cut} =  0.25 \rho_{max}$ is chosen to keep the initial magnetic field away from
the outer edge of the disk.  

The large-scale magnetic field is introduced in a very straightforward
manner by adding Wald's solution (W74) to the poloidal loops:
\begin{equation}
A_\phi = A_{\phi\,{\rm (loops)}} + 
{B_0 \over 2} \,\left(g_{\phi \phi} + 2\,a\,g_{t \phi}\right)
\end{equation}

The magnetic field is initialized 
using ${\cal{B}}^r =-\partial_\theta A_{\phi}$ and 
${\cal{B}}^\theta = \partial_r A_{\phi}$.  

\section{\bf Evolution Diagnostics\label{diags}}

The code diagnostics consist of shell-averaged and 
volume-integrated quantities computed at regular time intervals.  
The average of a quantity 
${\cal X}$ on a shell, radius $r$,
\begin{equation}\label{avgdef}
\langle{\cal X}\rangle(r,t) = {1 \over {{\cal A}}(r)} \int\int{ 
 {\cal X}\,\sqrt{-g}\, d \theta\,d \phi}
\end{equation}
where the area of a shell is ${\cal{A}}(r)$ and the bounds of integration 
range over the $\theta$ and $\phi$ grids.  
For these simulations, shell-averaged values of density,
$\langle\rho\rangle$, angular momentum,
$\langle \rho\, l\rangle$, gas pressure, $\langle P\rangle$,  
magnetic pressure $\langle{\|b\|}^2\rangle$/2, angular momentum
$\langle \rho\, h \, U_\phi\rangle$
and binding energy $\langle \rho\, h \, U_t\rangle$ are computed. 
Two sets of shell averages are available, based on whether the 
fluid at a given point on the shell is bound $(-h\,U_t \le 1)$ or unbound;
for instance,  the history of the unbound component of 
quantity ${\cal X}$ is written $\langle {\cal X}|_{-h\,U_t >1}\rangle(r,t)$. 

Fluxes through the shell are computed in the same manner, but are
not normalized with the area.  The rest mass flux
$\langle\rho\,U^r\rangle$, energy flux 
\begin{equation}
\langle{T^r}_t\rangle(r,t) = \langle{\rho\,h\,U^r\,U_{t}}\rangle(r,t)
+\langle{{\|b\|}^2\,U^r\,U_{t}}\rangle(r,t)-\langle{b^r\,b_t }\rangle(r,t),
\end{equation}
and the angular momentum flux
\begin{equation}
\langle{T^r}_{\phi}\rangle(r,t) = \langle{\rho\,h\,U^r\,U_{\phi}}\rangle(r,t)+
\langle{{\|b\|}^2\,U^r\,U_{\phi}}\rangle(r,t)-\langle{b^r\,b_\phi}\rangle(r,t)
\end{equation}
are computed. To ensure flexibility, each of the
three components in the above sums is stored in a separate file. In
addition, bound and unbound components of fluxes are computed and stored
separately.

Volume-integrated quantities are computed using
\begin{equation}\label{3avgdef}
\left[{\cal Q}\right](t) = \int\int\int{ 
 {\cal Q}\,\sqrt{-g}\, dr\,d \theta\,d \phi}.
\end{equation}
The volume-integrated quantities are the total rest mass, $\left[ \rho
U^t \right]$, angular momentum $\left[ {T^t}_{\phi}\right]$, and total
energy $\left[ {T^t}_{t}\right]$, distinguishing between bound and unbound material.


\end{document}